\documentclass[twocolumn]{aastex63}
\usepackage{hyperref}
\usepackage[colorinlistoftodos]{todonotes}
\usepackage{amsmath}
\usepackage{nicefrac}
\shorttitle{Bayesian inference of spin-orbit demographics}

\newcommand{\Princeton}{Department of Astrophysical Sciences, Princeton University, 4 Ivy
Lane, Princeton, NJ 08544, USA}

\begin{document}

\title{Ponderings on the Possible Preponderance of Perpendicular Planets}

\correspondingauthor{Jared Siegel}
\email{siegeljc@princeton.edu}

\author[0000-0002-9337-0902]{Jared C. Siegel}
\altaffiliation{NSF Graduate Research Fellow}
\affiliation{\Princeton}

\author[0000-0002-4265-047X]{Joshua N. Winn}
\affiliation{\Princeton}

\author[0000-0003-1762-8235]{Simon H. Albrecht}
\affil{Stellar Astrophysics Centre, Department of Physics and Astronomy, Aarhus University, Ny Munkegade 120, 8000 Aarhus C, Denmark}
\affiliation{\Princeton}
\begin{abstract}

Misalignments between planetary
orbits and the equatorial planes of their host
stars are clues about the formation and evolution of planetary systems.
Earlier work found evidence for a peak near $90^\circ$ in the
distribution of stellar obliquities,
based on frequentist tests.
We performed hierarchical Bayesian inference on a sample
of 174 planets for which either the full three-dimensional
stellar obliquity has been measured (72 planets) or for which
only the sky-projected stellar obliquity has been measured (102 planets).
We investigated whether the obliquities
are best described by a Rayleigh distribution, or by a mixture
of a Rayleigh distribution representing well-aligned systems
and a different distribution representing misaligned systems.
The mixture models are strongly favored over the single-component distribution.
For the misaligned component, we tried an isotropic distribution
and a distribution peaked at 90$^\circ$, and found the evidence
to be essentially the same for both models.
Thus, our Bayesian inference engine did not find strong evidence favoring a ``perpendicular peak,''
unlike the frequentist tests.
We also investigated selection biases that affect
the inferred obliquity distribution, such as the bias
of the gravity-darkening method against obliquities near $0^\circ$ or $180^\circ$.
Further progress in characterizing the obliquity distribution will
probably require the construction
of a more homogeneous and complete sample of measurements.\\
\\
\end{abstract}


\section{Introduction}
\label{sec:intro}

A planet's orbital motion is not necessarily aligned with its host star's rotation. The degree of misalignment
is quantified by the obliquity, $\psi$,
defined as the angle between the
direction $\hat{n}_{\star}$ of the star's rotational angular momentum and the direction $\hat{n}_\mathrm{o}$ of the planet's orbital angular momentum.
Measurements of the obliquity
or its sky-projection, $\lambda$, have been performed for
several hundred transiting exoplanets
\citep[see][for reviews]{Triaud2018, Albrecht2022}.

It is hoped that such measurements will
help us understand the processes
of planet formation and post-formation dynamical
interactions. For example, the observation that hot Jupiters
around low-mass stars tend to have low obliquities
has been interpreted as the outcome of tidal
dissipation within the star \citep{Winn2010},
while observations of high
obliquities have been variously attributed to
planet-planet scattering, Von Zeipel–Kozai–Lidov cycles,
and
other mechanisms \citep[see Section 4 of][]{Albrecht2022}.
In a few cases,
there are enough clues available to implicate
a particular obliquity-excitation mechanism;
for example, in the K2-290 system,
there is evidence that a companion
star was responsible for tilting the protoplanetary
disk \citep{Hjorth2021}.
Drawing population level conclusions will require knowledge of the
statistical distribution of stellar obliquities
and its dependence on the characteristics
of planetary systems and their host stars.

Modeling of the obliquity distribution has been
undertaken by \cite{Fabrycky2009},
\cite{Morton2014},
\cite{Mazeh2015},
\cite{Winn2017},
\cite{MunozPerets2018}, and
\cite{Louden2021},
using different techniques and
progressively larger samples.
In general, investigators have modeled
the distribution as a mixture of a
well-aligned component
and a very broad component, and have
shown evidence that the broad component
consists preferentially of ``hot'' stars
(with masses between 1.2 and 1.9~$M_\odot$
and effective temperatures between 6200 and 8000\,K).

Importantly, most of the measurement techniques are
not capable of determining $\psi$ directly,
but are instead
sensitive to either
the sky-projected obliquity ($\lambda$) or the difference between
the line-of-sight inclinations of the spin
and orbital axes ($i_\star$ and $i_{\rm o}$).
The various angles are related via the equation
\begin{equation}
    \label{eqn:obliquity}
    \hat{n}_{\star} \cdot \hat{n}_\mathrm{o}  = \cos \psi = \cos i_{\star} \cos i_\mathrm{o} + \sin i_{\star} \sin i_\mathrm{o} \cos \lambda.
\end{equation}

Recently, \cite{Albrecht2021} assembled a sample
of systems for which constraints on
$\lambda$, $i_{\rm o}$, and $i_\star$ are all
available, usually because more than one measurement
technique was applied to the same system.
In this sample, the misaligned component of
the obliquity distribution appeared to have a broad
peak at 80--125$^\circ$, rather than falling off
smoothly from $0^\circ$ or extending all the way
to 180$^\circ$.
Frequentist tests assigned low probabilities
($p\lesssim 10^{-3}$) to the hypothesis that
the peak is a statistical fluke, given the null hypothesis of isotropy.

The motivation for the work described in this paper
was to (i) repeat the frequentist tests now
that more data points have been obtained,
(ii) compare different models for the obliquity
distribution
using Bayesian hierarchical modeling,
and (iii) investigate selection effects that
might influence the results.
This paper is organized as follows.
Section~\ref{sec:sample} describes the available sample. Section~\ref{sec:frequentist} revisits the frequentist  tests of \cite{Albrecht2021}, and Section~\ref{sec:methods} outlines the Bayesian modeling framework. The results are presented in Section~\ref{sec:selection}.
In addition to analyzing the entire sample,
we examined subsamples that are expected
to be influenced by tides, or for which
tides should be negligible, as described
in Section~\ref{sec:sample_dependence}. Selection
biases are investigated in Section~\ref{sec:bias}.
Our conclusions are summarized
in Section~\ref{sec:concl}.

\section{Sample}
\label{sec:sample}

\begin{figure*}[t]
\gridline{\fig{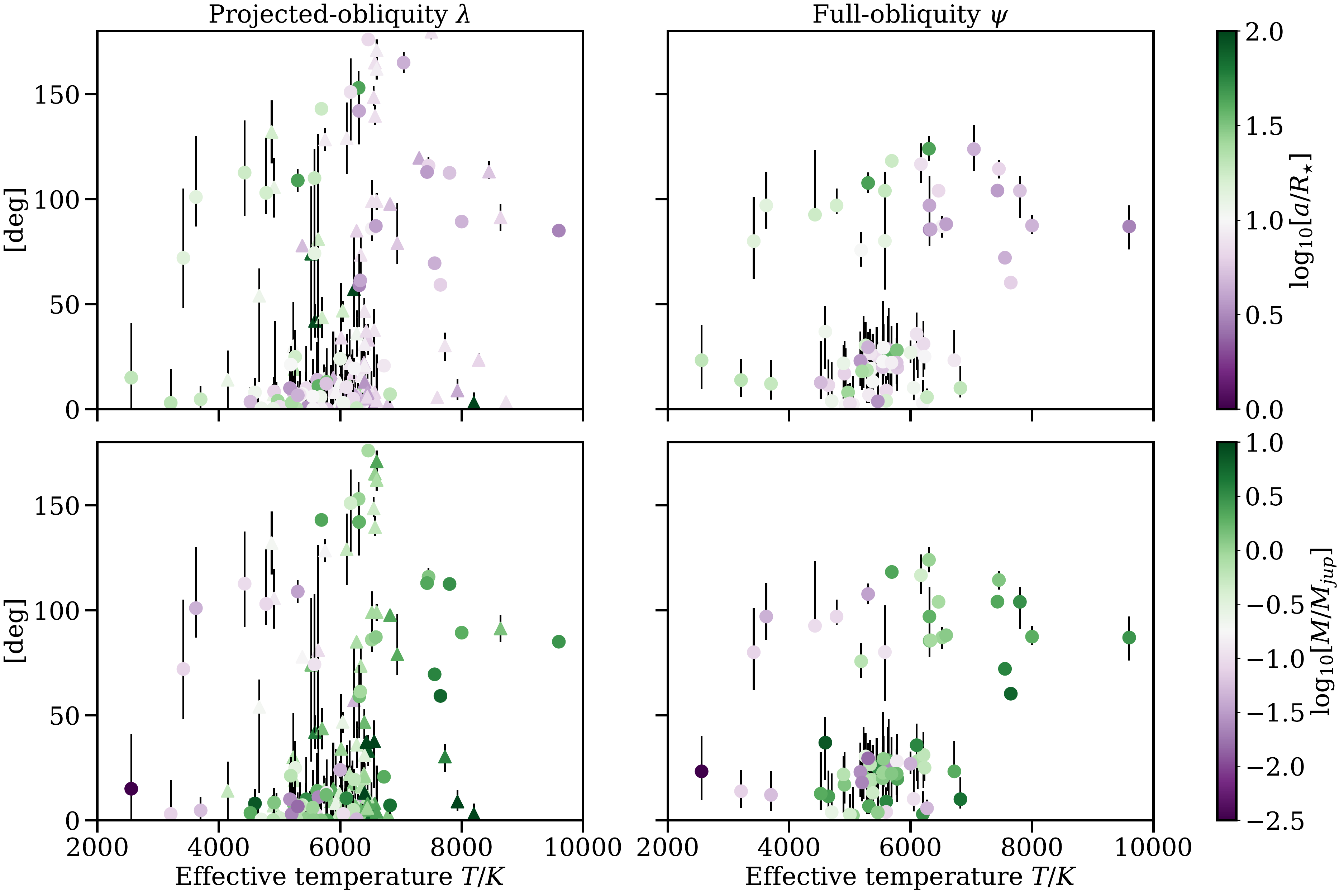}{
\textwidth}{ } }
\vskip -0.25in
\caption{Planetary and stellar parameters for systems with $\lambda$ measurements
and the subset of systems with $\psi$ measurements (right).
In the top and bottom rows, the color of the data
points conveys the planet's orbital separation and mass, respectively. Triangular data points are members
of the $\lambda$ sample, i.e.,
$\lambda$ has been measured but not
$\psi$. Circular data points are for
the $\psi$ sample, i.e.,
both $\lambda$ and $\psi$ have been measured.}
\label{fig:observed_sample_breakdown}
\end{figure*}

\cite{Albrecht2022} compiled a catalog of planets for
which measurements of the sky-projected obliquity
$\lambda$
are available, a subset of which also
have measurements of the full-obliquity $\psi$.
We supplemented this catalog with 16 newly available
measurements (Table~\ref{tab:sample_extra}),
of which 12 are for systems that previously
lacked any obliquity measurements, and 4 are for systems that now have full-obliquity information where only the projected-obliquity measurement was previously available.
The combined catalog consists of 174 planets.
Of these, 72 planets have measurements of both $\lambda$ and $i_\star$;
we will refer to these 72 planets as the $\psi$ sample.
For the other 102 planets, $\lambda$ has been
measured but there are no available
constraints on $i_\star$; we will refer to these
planets as the $\lambda$ sample.
Figure~\ref{fig:observed_sample_breakdown} displays
some key properties of the $\psi$ and $\lambda$ samples.

Most of the $\lambda$ measurements were obtained
by observing the Rossiter-McLaughlin effect,
and most of the $i_\star$ information 
comes from the combination of measurements of
projected rotation velocity $v\sin i$, rotation period,
and stellar radius \citep{Masuda2020}.
For KELT-9, KELT-17, Kepler-13, MASCARA-4, and WASP-189 the obliquity was inferred from the gravity-darkening method.
For HAT-P-7 and HD 89345, the stellar inclination was constrained via asteroseismology. 

\begin{deluxetable}{c | cccc}[t]
\tablecaption{Extension of the \cite{Albrecht2022} sample of
projected-obliquity $\lambda$ and obliquity $\psi$ measurements.\label{tab:sample_extra}}
\tablehead{
 \colhead{Name} & \colhead{$\lambda$ [deg]} &   \colhead{$\psi$ [deg]} &  \colhead{Update} &  \colhead{Ref.}
}
\startdata
55 Cnc e & $10.0_{-20.0}^{+17.0}$ & $23.0_{-12.0}^{+14.0}$ & --- & 1\\
HAT-P-3 b & $21.2\pm8.7$ & $75.7_{-7.9}^{+8.5}$ & \checkmark & 2\\
HAT-P-33 b & $5.9\pm4.1$ & --- & --- & 2\\
HAT-P-49 b & $97.7\pm1.8$ & --- & --- & 2\\
HD 3167 b & $6.6_{-7.9}^{+6.6}$ & $29.5_{-9.4}^{+7.2}$ & \checkmark & 3\\
HD 3167 c & $108.9_{-5.5}^{+5.4}$ & $107.7_{-4.9}^{+5.1}$ & \checkmark & 3\\
HD 89345 b & $74.2_{-32.5}^{+33.6}$ & $80.1_{-23.1}^{+22.3}$ & --- & 2\\
HIP 41378 d & $57.1_{-17.9}^{+26.4}$ & --- & --- & 4\\
K2-105 b & $81.0_{-47.0}^{+50.0}$ & --- & --- & 2\\
KELT-11 b & $77.86_{-2.26}^{+2.36}$ & --- & --- & 5\\
Qatar-6 A b & $0.1\pm2.6$ & $21.82_{-18.36}^{+8.86}$ & --- & 6\\
TOI-1478 b & $6.2_{-5.5}^{+5.9}$ & --- & --- & 7\\
TOI-2076 b & $3.0_{-15.0}^{+16.0}$ & $18.0_{-9.0}^{+10.0}$ & --- & 8\\
TOI-640 b & $176.0\pm3.0$ & $104.0\pm2.0$ & --- & 9\\
WASP-156 b & $105.7_{-14.4}^{+14.0}$ & --- & --- & 2\\
WASP-47 b & $0.0\pm24.0$ & $29.2_{-13.3}^{+11.1}$ & \checkmark & 2
\enddata
\tablecomments{A check mark in the ``Update''
column indicates a planet that appeared
in the \cite{Albrecht2022} catalog
for which updated measurements are now
available. References:
1 \citep{Zhao2022},
2 \citep{Bourrier2023},
3 \citep{Bourrier2021},
4 \citep{Grouffal2022},
5 \citep{Mounzer2022},
6 \citep{Rice2023},
7 \citep{Rice2022},
8 \citep{Frazier2022},
9 \citep{Knudstrup2023}. }
\end{deluxetable}

\section{Frequentist Hypothesis Testing}
\label{sec:frequentist}

\cite{Albrecht2021} noted an
apparent concentration of obliquities
near $90^\circ$ in the
$\psi$ sample that
was available at the time.
Focusing attention on the
19 planets with $\cos\psi< 0.75$ in their sample, they asked: under the null hypothesis that the spin and orbital axes
are uncorrelated, how often would the distribution of $\psi$
be concentrated near $90^\circ$ as strongly as seen in the real data?
The concentration was quantified by either the dispersion of $\cos \psi$ around $0$ or by the standard deviation of $\cos \psi$,
and in both cases the null hypothesis was rejected
with $p\lesssim 10^{-3}$.

In the updated sample, 25 systems satisfy
the misalignment criterion $\cos\psi < 0.75$.
In each of $10^6$ trials, we
drew 25 values 
of $\cos \psi$ from the distribution
$\mathcal{U}(-1, 0.75)$ representing the null hypothesis.
We then drew values 
of $\cos \psi$ from the measurement posteriors of the 25 systems for which $\cos\psi< 0.75$. 
Unlike \cite{Albrecht2021}, who approximated the measurement posteriors as skew-normal distributions, we drew directly from the measurement posteriors.
Using this method and the same sample as \cite{Albrecht2021},
the null hypothesis was rejected with median p--values of $p=2.5 \times 10^{-3}$ and $0.8 \times 10^{-3}$, in terms of the dispersion and standard deviation respectively. 
After adding the new and updated measurements,
the null hypothesis was rejected with median p--values of $p=0.5 \times 10^{-3}$ and $0.2 \times 10^{-3}$.
Thus, regardless of whether the concentration was quantified by the dispersion or standard deviation, the null hypothesis sample was rejected with high confidence,
and the new data reduced the corresponding $p$-values by factors of
5 and 4.

As acknowledged by \cite{Albrecht2021}, these
frequentist tests are difficult to interpret because
the tests were
devised after seeing the data.
Also, frequentist tests
can demonstrate
that the data are inconsistent with the
null hypothesis, but they leave one
wondering which probability
distributions {\it are} consistent with the data. Answering the latter
question was the goal
of the hierarchical Bayesian modeling
described below.

\section{Hierarchical Bayesian Inference}
\label{sec:methods}

\subsection{Formalism}
\label{sec:formalism}

For a given dataset $\vec{d}$,
a property of the system $x$
can be inferred from the posterior probability distribution
\begin{equation}
     p(x \mid \vec{d}) \propto p( \vec{d} \mid x) ~\pi_0(x),
\end{equation}
where $p( \vec{d} \mid x)$ is the likelihood function and $\pi_0(x)$ is the prior on $x$.
For example,
$\vec{d}$ might represent
a time
series of radial velocities exhibiting the Rossiter-McLaughlin
effect, and $x$ might represent $\lambda$.

In a hierarchical model, we use
a collection of $N$ datasets $ \{ \vec{d}_1, \dots, \vec{d}_N \}$
to infer the parameters $\Theta$ of a model for the
distribution $f_x (x \mid \Theta)$.
Assuming the datasets $ \{ \vec{d}_i \}$ are independent of each other, the likelihood function for the model parameters is 
\begin{align}
    \label{eqn:heir_orig}
    \mathcal{L}(\{ \vec{d}_i \} \mid  \Theta) &= \prod_{i=1}^N \int p( \vec{d}_i \mid x ) f_x(x \mid \Theta) \,dx,\\
    &\propto \prod_{i=1}^N \int p( x  \mid \vec{d}_i ) \frac{ f_x(x \mid \Theta)}{ \pi_0(x)} dx.
\end{align}
The population distribution $f_x (x \mid \Theta)$ is essentially treated as a prior on $x$; marginalizing over $x$ then leaves a likelihood function for $\Theta$. 
Throughout this work, the likelihood function for the model parameters is denoted $\mathcal{L}(\mathrm{data} \mid \mathrm{model})$;  all other probability density functions are denoted $p(X \mid Y)$.

For our application,
we modeled the obliquity distribution
$f_{\cos \psi}( \cos \psi \mid \Theta)$
using different samples of planets, and different likelihood
functions for each sample.
The choices for the likelihood function were
\begin{itemize}
\item The $\lambda$ likelihood,
based only on measurements
of $\lambda$, without any constraints on $i_\star$.
\item The $\psi$ likelihood,
based on measurements of $\psi$, or
separate measurements of both $\lambda$ and $i_\star$.
\item The $\psi\,\&\,\lambda$
likelihood, i.e., the combination
of $\psi$ likelihoods for systems
where constraints on $\psi$
are available, and
$\lambda$ likelihoods
for systems for which
only $\lambda$ has been
measured.
\end{itemize}

For the $\lambda$ likelihood,
the data set consists of $N_\lambda$ posteriors
for $\lambda$, and
we relate $\lambda$ and $\psi$ using
\begin{equation}
    \tan \lambda = \tan \psi \sin \phi,
\end{equation}
an equation valid for $i_{\rm o} \approx 90^\circ$,
as is the case for transiting planets
considered throughout this work.
Here, $\phi$ is the azimuthal angle of $\hat{n}_\star$ in a spherical polar coordinate system in which $\hat{n}_{\rm o}$ is the polar axis. The
azimuthal angle is assumed to be uniformly distributed, i.e.,
\begin{equation}
    f_{ \sin \phi }(x) = \frac{2}{\pi} \frac{1}{ \sqrt{1-x^2} }.
\end{equation}
The likelihood function for $\Theta$
is
\begin{align}
    \label{eqn:like_lambda}
    \mathcal{L}_\lambda(   \{ \vec{d}_i \} \mid \Theta ) \propto 
    \prod_{i=1}^{N_\lambda}  & \int_{ \cos \psi, \sin \phi} d \cos \psi \times d \sin \phi\nonumber \\
    & \times p_{\lambda} (\tan^{-1}[ \tan \psi   \sin \phi] \mid \vec{d}_i) \nonumber \\
    & \times f_{\cos \psi}(\cos \psi \mid \Theta)  \nonumber \\
    & \times f_{\sin \phi}(\sin \phi),
\end{align}
where $p_{\lambda}$ is the likelihood function for $\lambda$ given $\vec{d}_i$,
the data for the $i$th system. 
This likelihood function is made computationally tractable via the $K$-samples approximation \citep{Hogg2010}:
\begin{align}
    \mathcal{L}_\lambda( & \{ \vec{d}_i \} \mid  \Theta )\propto \nonumber \\ 
     & \prod_{i=1}^{N_\lambda} \frac{1}{K} \sum_{k=1}^{K}  \mathcal{N} ( \tan^{-1}[ \tan \psi_{k,\Theta} \sin \phi_k] \mid \lambda_i, \sigma_{\lambda_i} ),
\end{align}
where $\cos \psi_{k,\Theta}$ is the $k$th draw from $f_{\cos \psi}(\cos \psi \mid \Theta)$, $\sin \phi_k$ is the $k$th draw from $f_{ \sin \phi}$, and $K=1000$.
For inference of the underlying obliquity distribution, we
found it convenient to sample from the distribution $f_{\cos \psi}$ and approximate each measurement posterior as a normal distribution.

For the $\psi$ likelihood, the details
depend on the measurement technique.
In most cases,
there are measurement posteriors for $\lambda$, $v \sin i$, $P_\mathrm{rot}$, and $R_{\star}$,
and the stellar inclination is constrained using the equation
\begin{equation}
\label{eq:vsini}
    \sin i = \frac{P_\mathrm{rot}}{2 \pi R}~v \sin i,
\end{equation}
from which the obliquity can be calculated
using Equation~\ref{eqn:obliquity}.
The likelihood function in this case is
\begin{align}
    \mathcal{L}_\psi(   \{ \vec{d}_i \} \mid \Theta ) \propto 
    \prod_{i=1}^{N_\psi}  & \int_{ \cos \psi, \sin \phi, v} d \cos \psi \times d \sin \phi \times d v \nonumber \\
    & \times p_{v \sin i} (v  \cos \psi / \cos \lambda \mid \vec{d}_i) \nonumber \\
    & \times p_{\lambda} (\tan^{-1}[ \tan \psi  \sin \phi] \mid \vec{d}_i) \nonumber \\
    & \times p_{v}(v\mid \vec{d}_i) \nonumber \\
    & \times f_{\cos \psi}(\cos \psi \mid \Theta)  \nonumber \\
    & \times f_{\sin \phi}(\sin \phi).
\end{align}
For notational convenience, we combined the $P_{\mathrm{rot},i}$ and $R_{\star,i}$ measurements into a likelihood for the rotation velocity $p_v(v \mid \vec{d}_i)$.
We evaluated the likelihood function via the $K$-samples approximation, drawing from the $f_{\cos \psi}$, $f_{\sin \phi}$, and $p_v$ distributions.

In the cases for which the obliquity
determination
was based on asteroseismology or gravity darkening,
we used the measurement posterior for $\cos \psi$,
giving a likelihood function
\begin{align}
    \mathcal{L}_\psi(   \{ \vec{d}_i \} \mid \Theta ) \propto 
    \prod_{i=1}^{N_\psi}  & \int_{ \cos \psi} d \cos \psi \nonumber \\
    & \times p_{ \cos \psi} ( \cos \psi \mid \vec{d}_i) \nonumber \\
    &\times f_{\cos \psi}(\cos \psi \mid \Theta).
\end{align}
For our $K$-samples approximation, we approximated the
posterior for $\cos\psi$ as a normal distribution.

\subsection{Models for the obliquity distribution}

\begin{figure}[t]
\gridline{\fig{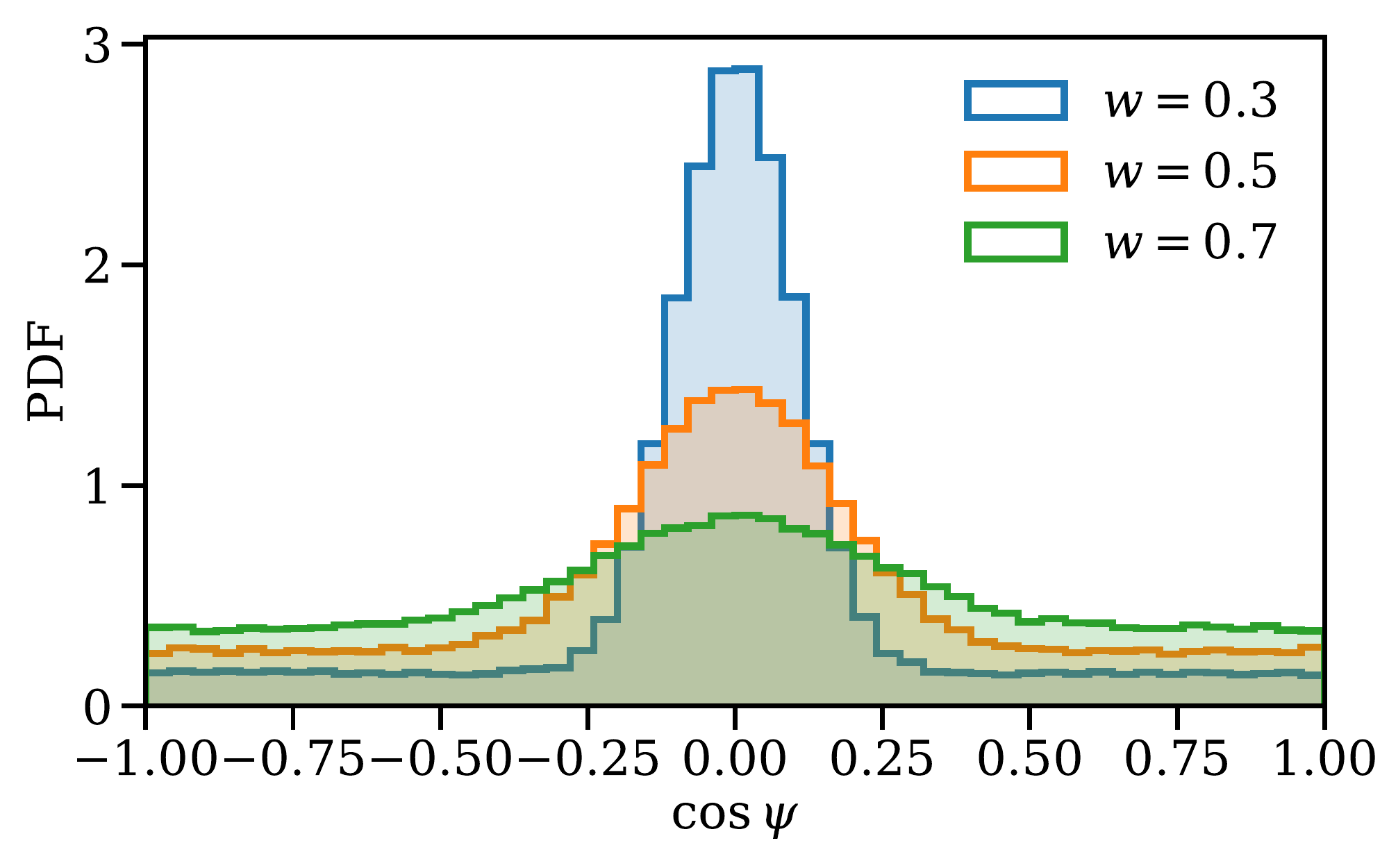}{\columnwidth}{} }
\caption{ The isonormal
distribution used in Model~2, for different
choices of $w$. }
\label{fig:IsoGauss}
\end{figure}

Four models for the
obliquity distribution
were considered. The models were designed to determine whether the obliquity distribution has multiple components,
and if so, whether the misaligned component is peaked at a particular value.
In order of increasing complexity, the models are:
\begin{enumerate}
\setcounter{enumi}{-1}

\item A Rayleigh distribution for $\psi$ with a single parameter $\sigma_R$ specifying the
mode and width of the distribution.

\item A mixture of a Rayleigh distribution for $\psi$
and an isotropic distribution
(uniform in $\cos\psi$).
In addition
to $\sigma_R$, this model has a parameter $q$ specifying the fraction
of systems that are drawn from the Rayleigh distribution.

\item A mixture of a Rayleigh distribution for $\psi$
and an ``isonormal'' distribution,
a function of our devising that interpolates between an isotropic distribution
and a normal distribution (see~Figure~\ref{fig:IsoGauss}):
\begin{align}
  f_\mathrm{isonormal}(\cos \psi \mid \mu, w) \propto & ~w f_\mathrm{isotropic}(\cos \psi) \nonumber \\
  +& (1-w)\,\mathcal{N}\left(\cos \psi \mid \mu, \frac{w}{3} \right),
\end{align}
where $\mathcal{N}$ is a normal
distribution with mean $\mu$ and
standard deviation $\nicefrac{w}{3}$.
As $w \rightarrow 0$, $f_\mathrm{IG}$ tends toward a Dirac delta
function centered at $\mu$, and 
as $w \rightarrow 1$, $f_\mathrm{isonormal} \rightarrow f_\mathrm{isotropic}$. 
For this model, $\mu$ is held fixed at $0$ (the ``perpendicular preponderance'') and the adjustable parameters are $\sigma_R$, $q$, and $w$.

\item The same as Model 3 but allowing the mean $\mu$ of the peaked
distribution to be an adjustable parameter, for
a total of four parameters: $\{\sigma_R, q, \mu, w\}$.

\end{enumerate}
These models were selected because they
have a minimal number of free parameters
and are easy to interpret.
Each model is normalized such that $\int_{-1}^1 f_{\cos \psi} d \cos \psi =1$. 
We adopted the following priors on the model parameters:
\begin{align*}
    \sigma_R &\sim \begin{cases}
      \mathcal{U}(0.3, 2.0~\mathrm{rad})~\mathrm{in~Model~0},\\
      \mathcal{U}(\frac{\pi}{180}, 0.5~\mathrm{rad})~\mathrm{in~Models~1\mbox{--}3},
\end{cases}\\
    q &\sim \mathcal{U}(0,1),\\
    w &\sim \mathcal{U}(0,1),\\
    \mu &\sim \mathcal{U}(-0.5, 0.5).
\end{align*}
The lower bounds on $\sigma_R$ were introduced to
avoid numerical instabilities; as long as they are sufficiently small,
the specific values do
not materially affect the results.

\begin{deluxetable*}{ccc | r@{}c@{}l r@{}c@{}l r@{}c@{}l }[t]
\tablecaption{Bayes Factors $R$ relative to Model 1\label{tab:R_all}}
\tablehead{
 \colhead{Restriction} & \colhead{Sample} &
 \colhead{$N$} & \multicolumn{3}{c}{Model 0} & \multicolumn{3}{c}{Model 2} & \multicolumn{3}{c}{Model 3}
}
\startdata
None  & $\psi$ & 72 & $(14 \pm 8)$& $\times$ & $10^{-20}$ & $2.7 $ & $\pm$ & $ 0.5$ & $1.2 $ & $\pm$ & $ 0.2$\\
 & $\lambda$ & 102 & $(5 \pm 2)$& $\times$ & $10^{-14}$ & $0.382 $ & $\pm$ & $ 0.007$ & $0.307 $ & $\pm$ & $ 0.006$\\
 & $\psi$ \& $\lambda$ & 174 & $(3 \pm 2)$& $\times$ & $10^{-29}$ & $1.1 $ & $\pm$ & $ 0.8$ & $0.6 $ & $\pm$ & $ 0.2$\\
 \hline
No G.D.  & $\psi$ & 67 & $(8 \pm 6)$& $\times$ & $10^{-20}$ & $1.6 $ & $\pm$ & $ 0.1$ & $0.85 $ & $\pm$ & $ 0.08$\\
 & $\lambda$ & 102 & $(37 \pm 3)$& $\times$ & $10^{-15}$ & $0.377 $ & $\pm$ & $ 0.008$ & $0.308 $ & $\pm$ & $ 0.006$\\
 & $\psi$ \& $\lambda$ & 169 & $(3 \pm 4)$& $\times$ & $10^{-30}$ & $0.5 $ & $\pm$ & $ 0.1$ & $0.4 $ & $\pm$ & $ 0.1$\\
 \hline
No G.D. \& Giants with Weak Tides  & $\psi$ & 31 & $(34 \pm 1)$& $\times$ & $10^{-8}$ & $1.08 $ & $\pm$ & $ 0.03$ & $0.69 $ & $\pm$ & $ 0.02$\\
 & $\lambda$ & 70 & $(42 \pm 2)$& $\times$ & $10^{-10}$ & $0.74 $ & $\pm$ & $ 0.01$ & $0.434 $ & $\pm$ & $ 0.006$\\
 & $\psi$ \& $\lambda$ & 101 & $(76 \pm 10)$& $\times$ & $10^{-15}$ & $0.84 $ & $\pm$ & $ 0.03$ & $0.53 $ & $\pm$ & $ 0.02$
\enddata
\tablecomments{G.D. is gravity darkening. $R$ is the ratio of the models' posterior probabilities, given the data and
assuming equal prior probabilities.
$R>1$ implies $M_i$ is favored over $M_1$ with $R : 1$ odds.
}
\end{deluxetable*}

\section{Bayesian Model Selection }
\label{sec:selection}

The Bayes factor (or evidence ratio) was used
to evaluate the competing models.
For a data set $\{ \vec{d}_i \}$ and model $M$ with free-parameters $\Theta$, the Bayesian evidence is defined as
\begin{equation}
    p( \{ \vec{d}_i \} \mid M) = \int \mathcal{L}(  \{ \vec{d}_i \} \mid M, \Theta) \pi_0(\Theta)  d \Theta,
\end{equation}
where $\mathcal{L}(  \{ \vec{d}_i \} \mid M, \Theta)$ is the likelihood function and $\pi_0(\Theta)$ is the prior on the model parameters. The Bayesian evidence for each model was calculated via the $K$-samples approximation, using a number
of samples equal to the number of free parameters times $10^6$.
The sampling error in the calculation of the Bayesian evidence was estimated by redrawing from the Monte Carlo samples 100 times for each model.

Given two competing models $M_0$ and $M_1$, the Bayes factor is
\begin{equation}
    R = \frac{ p( \{ \vec{d}_i \} \mid M_0) }{ p( \{ \vec{d}_i \} \mid M_1)  }.
\end{equation}
If the prior probabilities of models $M_0$ and $M_1$ are equal, $R$ is
the ratio of the posterior probabilities for $M_0$ and $M_1$ given the data $\{ \vec{d}_i \}$. In this case, a Bayes factor of $R>1$ implies $M_0$ is favored over $M_1$ with $R : 1$ odds. Bayes factors of $R=10,$ $20,$ and $100$ correspond to odds ratios of $91,$ $95,$ and $99\%$, respectively; see \cite{Trotta2007} and \cite{Abbott2018} for reviews and applications of the Bayes factor.

\subsection{Full-obliquity modeling}
\label{sec:obliquity_only}

\begin{deluxetable}{cc | ccc}[t]
\tablecaption{Parameters of the model
obliquity distributions, based on complete measurement samples.\label{tab:par_all}}
\tablehead{
\colhead{Sample} & & \colhead{$\psi$}  & \colhead{$\lambda$} & \colhead{$\psi\,\&\,\lambda$}\\ 
 \colhead{N} &  & \colhead{72} & \colhead{102} & \colhead{174}}
\startdata
 Model 0 & $\sigma_R$ [deg] & $44^{+3}_{-3}$ & $42^{+3}_{-3}$ & $42^{+3}_{-3}$\\
\hline
 Model 1 & $\sigma_R$ [deg] & $2.0^{+0.9}_{-0.6}$ & $9^{+3}_{-2}$ & $5^{+1}_{-1}$\\
 & $q$ & $0.55^{+0.05}_{-0.05}$ & $0.65^{+0.06}_{-0.06}$ & $0.58^{+0.04}_{-0.03}$\\
\hline
 Model 2 & $\sigma_R$ [deg] & $2.1^{+1.0}_{-0.6}$ & $10^{+3}_{-3}$ & $5^{+1}_{-1}$\\
 & $q$ & $0.58^{+0.05}_{-0.06}$ & $0.67^{+0.06}_{-0.06}$ & $0.6^{+0.04}_{-0.04}$\\
 & $w$ & $0.7^{+0.2}_{-0.1}$ & $0.8^{+0.1}_{-0.2}$ & $0.8^{+0.1}_{-0.1}$\\
\hline
 Model 3 & $\sigma_R$ [deg] & $2.1^{+1.0}_{-0.6}$ & $10^{+3}_{-3}$ & $5^{+2}_{-1}$\\
 & $q$ & $0.6^{+0.06}_{-0.06}$ & $0.66^{+0.06}_{-0.07}$ & $0.59^{+0.06}_{-0.04}$\\
 & $w$ & $0.7^{+0.2}_{-0.2}$ & $0.8^{+0.1}_{-0.2}$ & $0.83^{+0.09}_{-0.12}$\\
 & $\mu$ & $0.0^{+0.2}_{-0.2}$ & $0.0^{+0.3}_{-0.3}$ & $0.0^{+0.3}_{-0.2}$
\enddata
\tablecomments{ Parameter posteriors for each combination of model (e.g., 0, 1, 2, or 3) and type of data ($\psi$ collection, $\lambda$ collection, and the $\psi \& \lambda$ collection). The reported posteriors are the 16, 50, and 84th percentiles. }
\end{deluxetable}

\begin{deluxetable}{cc | ccc}[t]
\tablecaption{Parameters of the model
obliquity distributions, based on giant planets
with weak tides and removing the gravity darkening measurements.\label{tab:par_decoupled}}
\tablehead{
\colhead{Sample} & & \colhead{$\psi$}  & \colhead{$\lambda$} & \colhead{$\psi\,\&\,\lambda$}\\ 
 \colhead{N} &  & \colhead{31} & \colhead{70} & \colhead{101}}
\startdata
 Model 0 & $\sigma_R$ [deg] & $50^{+6}_{-5}$ & $40^{+3}_{-3}$ & $43^{+3}_{-3}$\\
\hline
 Model 1 & $\sigma_R$ [deg] & $2.1^{+1.7}_{-0.8}$ & $13^{+3}_{-3}$ & $8^{+2}_{-2}$\\
 & $q$ & $0.43^{+0.08}_{-0.08}$ & $0.69^{+0.07}_{-0.08}$ & $0.56^{+0.06}_{-0.06}$\\
\hline
 Model 2 & $\sigma_R$ [deg] & $2.2^{+2.1}_{-0.9}$ & $14^{+3}_{-3}$ & $9^{+3}_{-2}$\\
 & $q$ & $0.45^{+0.08}_{-0.08}$ & $0.73^{+0.07}_{-0.07}$ & $0.59^{+0.06}_{-0.06}$\\
 & $w$ & $0.7^{+0.2}_{-0.2}$ & $0.7^{+0.2}_{-0.2}$ & $0.7^{+0.2}_{-0.2}$\\
\hline
 Model 3 & $\sigma_R$ [deg] & $2.1^{+2.0}_{-0.8}$ & $14^{+3}_{-3}$ & $8^{+3}_{-2}$\\
 & $q$ & $0.45^{+0.08}_{-0.08}$ & $0.71^{+0.07}_{-0.08}$ & $0.58^{+0.06}_{-0.06}$\\
 & $w$ & $0.7^{+0.2}_{-0.2}$ & $0.8^{+0.2}_{-0.2}$ & $0.8^{+0.1}_{-0.2}$\\
 & $\mu$ & $0.1^{+0.2}_{-0.3}$ & $0.0^{+0.3}_{-0.2}$ & $0.1^{+0.2}_{-0.2}$
\enddata
\tablecomments{ Parameter posteriors for each combination of model (e.g., 0, 1, 2, or 3) and type of data ($\psi$ collection, $\lambda$ collection, and the $\psi \& \lambda$ collection). The reported posteriors are the 16, 50, and 84th percentiles. }
\end{deluxetable}

To begin, the evidence and the posteriors for
the parameters of each obliquity distribution model
were inferred using the $\psi$ sample, i.e., the
72 planets for which information
about both $\lambda$ and $i_\star$
is available.
Here and in the other cases
described below, the posteriors for the model parameters were inferred with a Markov Chain Monte Carlo method,
using 40 walkers for 50{,}000 iterations, of which
the first third were discarded as burn-in (a conservative threshold based on visual convergence of the walker chains).
Table~\ref{tab:R_all} gives the Bayes factors relative to Model 1,
and Table~\ref{tab:par_all} gives the 16, 50, and 84th percentiles of the
posteriors of the model parameters.

\begin{figure*}[t]
\gridline{\fig{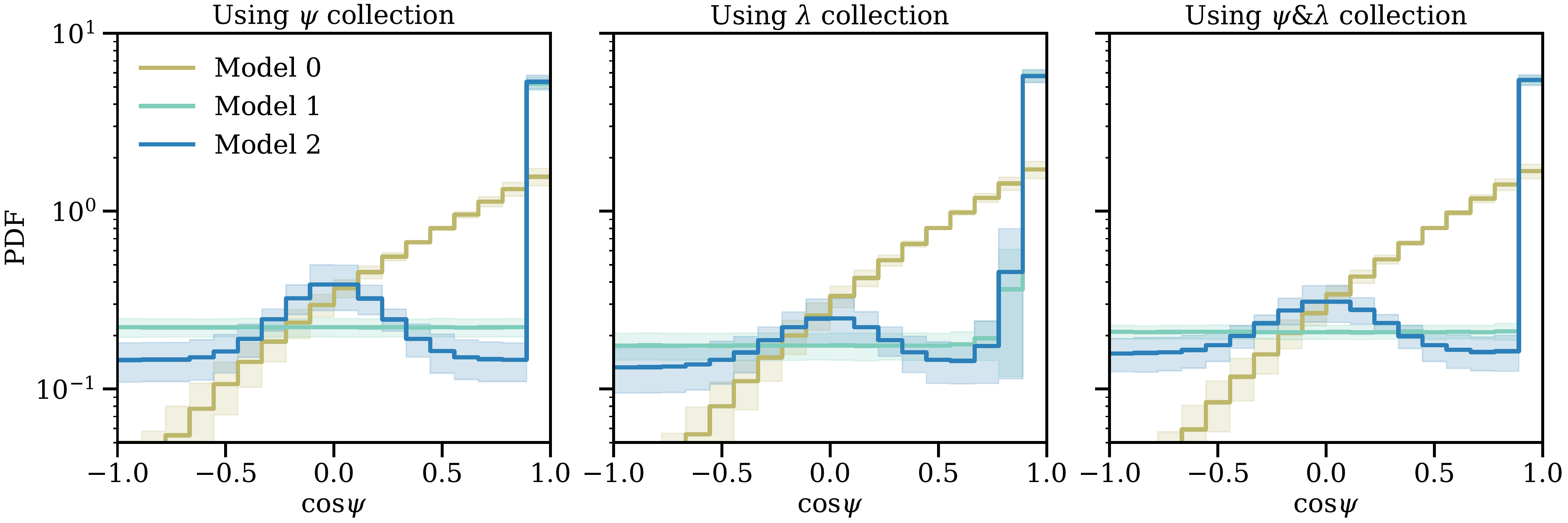}{\textwidth}{} }
\vskip -0.25in
\caption{Inferred obliquity
distributions for different
samples and likelihoods.
Regardless of the sample, the peak in the isonormal
 model is small enough to be consistent or nearly consistent
 with the isotropic model.
Left: The $\psi$ sample of 72 planets
for which measurements of both
$\lambda$ and $i_\star$ are
available, analyzed with the $\psi$ likelihoods.
Center: The $\lambda$ sample of 102 planets,
analyzed with the $\lambda$ likelihoods.
Right: The entire $\psi\,\&\,\lambda$ sample of 174 planets,
analyzed with the $\psi\,\&\,\lambda$ likelihoods.
Solid lines demarcate the mean, and the shaded regions show the $1\sigma$ uncertainty intervals.
}
\label{fig:comparing_models}
\end{figure*}

In Model 0 (Rayleigh), the width $\sigma_R$ was found to be
$44^{\circ}\pm 3^\circ$.
This model was strongly disfavored ($R < 10^{-9}$) relative
to the other models in which the misaligned systems were
described by a separate component.
We conclude that the obliquity distribution is best
described as a mixture between a well-aligned component and a misaligned component, rather than a single smooth distribution.

In the multi-component models (1--3), the fraction of systems
belonging to the Rayleigh component was $\approx$60\%,
and $\sigma_R$ was found to be
$\approx$2$^\circ$.
Regarding the misaligned components, the Bayesian evidence favors the peaked
models (2 and 3) over the isotropic distribution (1), but not significantly.
Model~2 (Rayleigh~$+$~90$^\circ$ peak)
is favored over Model~1 (Rayleigh~$+$~isotropic)
with $R\approx 3$.
Model~3, in which the peak location is allowed to vary,
is consistent with a peak at 90$^{\circ}$; 
however, the constraint on $\mu$ is weak and the
degree of concentration is low.
For Models~2--3, the width parameter $w$ was found to be $\gtrsim$0.7,
indicating a broad distribution.
The inferred obliquity distributions are visualized in Figure~\ref{fig:comparing_models}. 

For reference, we repeated the inference using the same sample
of 72 planets, but using
only the $\lambda$ likelihoods and ignoring
the constraints on the stellar
inclination angle.
The results are nearly the same as those that were
obtained using the $\psi$ likelihoods, all consistent within $1\sigma$. Evidently, the inclination information has only a minor effect on the results.\footnote{This
conclusion was also reached
by Dong \& Foreman-Mackey in work
shared with us prior to publication.}

\subsection{Projected-obliquity modeling}
\label{sec:lambda_only}

Next, we considered the $\lambda$ sample,
consisting of 102 planets for which only $\lambda$ was measured and no information is available
about the stellar inclination.
Table~\ref{tab:R_all} gives the Bayes factors relative to Model 1,
and Table~\ref{tab:par_all} gives the 16, 50, and 84th percentiles of the model posteriors.
The inferred obliquity distributions are visualized in 
Figure~\ref{fig:comparing_models}. 

Again, Model 0 (Rayleigh) was strongly disfavored relative to the
multi-component models.
Also similar to the previous case is that
Models 1--3 were found to be effectively indistinguishable
by the Bayesian evidence; Model 1 is
favored with odds ratios $\approx$3.
In Models 2--3, the
width parameter $w$ was found to be $0.8_{-0.2}^{+0.1}$.
Relative to the $\psi$ sample, the $\lambda$ sample favors a larger width for the well-aligned component ($\sigma_R \approx 10^{\circ}$ compared to $\sigma_R \approx 2^{\circ}$) and provides weaker evidence for models with a peaked
misaligned component;
these trends may indicate underlying differences between the $\psi$ and $\lambda$ sample, which we investigate in Section~\ref{sec:bias}. 

\subsection{Joint obliquity and projected-obliquity modeling}
\label{sec:obliquity_and_lambda}

Finally, we took advantage of all the available information.
We combined the collection of $\psi$ constraints from
the 72 systems in the $\psi$ sample
with the $\lambda$ constraints from the 102 systems
in the $\lambda$ sample.
The Bayes factors relative to Model 1 are presented in Table~\ref{tab:R_all}. Table~\ref{tab:par_all} presents the 16, 50, and 84th percentiles of the model posteriors. The inferred obliquity distributions are visualized in Figure~\ref{fig:comparing_models}.
Model 0 was strongly disfavored relative to Models 1--3,
and Models 1--3 were effectively indistinguishable.
The preferred models are a mixture of a narrow Rayleigh distribution and
a very broad or isotropic component.

To summarize, the single-component Model 0 was always trounced
by the multi-component Models 1--3.
While the results of analyzing the $\psi$ and $\lambda$ systems differ in detail, neither the isotropic model nor the peaked
models were strongly favored
by the Bayesian evidence.

\begin{figure*}[t]
\gridline{\fig{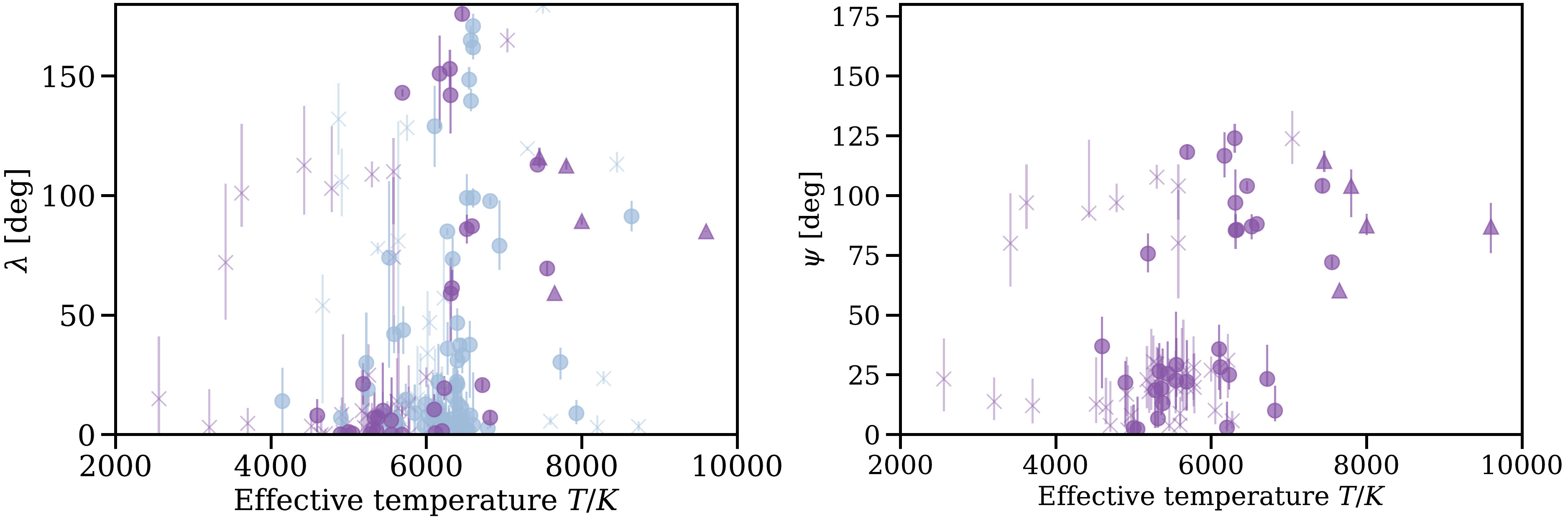}{
\textwidth}{ } }
\vskip -0.25in
\caption{ 
Projected-obliquity (left) and full-obliquity (right) versus stellar effective temperature. 
Blue points
represent planets for which $\lambda$ has been measured but not $\psi$. 
Purple points represent
planets for which $\psi$ has been measured.
Circles
are the giant-planet hosts in the weak-tides sample,
triangles are the gravity-darkening measurements, and crosses are all other measurements.}
\label{fig:decoupled_sample_selection}
\end{figure*}

\section{Sample dependence}
\label{sec:sample_dependence}

The sample of planets covers a wide range of orbital separations, planet masses, and stellar types. The resulting diversity of possible modes of planet formation and evolution makes the obliquity distribution of the entire sample difficult
to interpret. In particular, for massive close-orbiting planets, dissipative tidal interactions generally lead to orbital circularization, synchronization of spin and orbital
periods, and spin-orbit alignment. Tidal dissipation is expected to be faster
for low-mass host stars with thick convective envelopes (with effective temperatures
below the ``Kraft break''
at about 6250\,K) than for more massive stars with radiative envelopes.
Although the mechanism by which tidal dissipation occurs
is uncertain, the obliquity data shows the qualitative trends
one would expect from tidal dissipation.
Close-orbiting giant planets around low-mass stars generally have
low obliquities, while a broader distribution of obliquities is seen
in systems with smaller planets, wider orbits, or hotter stars \citep{Winn2010, Albrecht2012, Albrecht2022}. 

In an effort to simplify the interpretation of the obliquity distribution,
we repeated our Bayesian model comparisons
after applying restrictions to the samples
intended to group together
systems where one might expect similar formation mechanisms
and obliquity-altering mechanisms to have occurred.
We focused on giant planets ($M_p > 0.3 M_{\rm Jup})$, and included those with
$a / R_{\star} < 300$ (a criterion that
excluded
only $\beta$~Pic\,b).
We omitted the 5 systems for which
the obliquity was determined
with the gravity darkening method,
out of concern about a
selection bias favoring
large misalignments
(see Section~\ref{sec:GD_bias}).
For a system to be categorized as having ``weak tides'',
we required $a / R_{\star} > 7$
or $T_\mathrm{\rm eff} < 6250$~K.
Figure~\ref{fig:decoupled_sample_selection} shows key properties of the samples and identifies the weak-tides
systems.

\begin{figure*}[t]
\gridline{\fig{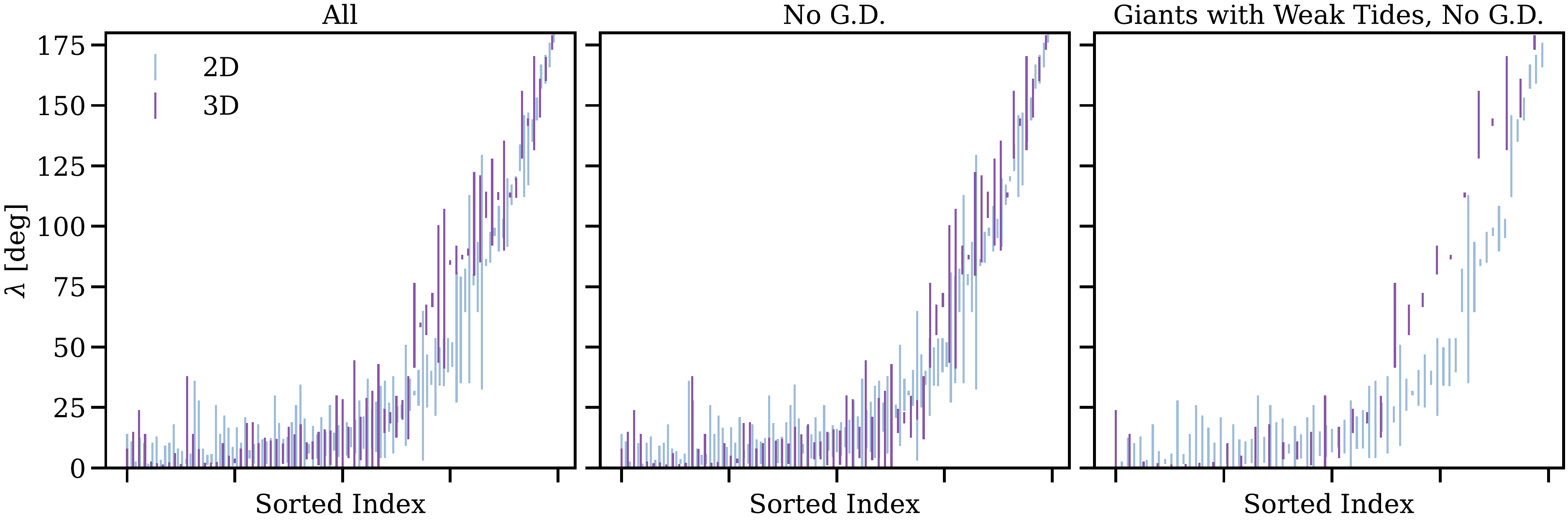}{
\textwidth}{ } }
\vskip -0.25in
\caption{ The $\lambda$ measurements from the $\psi$ sample
are
slightly but systematically higher than the $\lambda$
measurements from the $\lambda$ sample.
This effect is likely due
at least in part to
the selection bias of the gravity darkening method,
which was used in 5 cases in the $\psi$ sample
and no cases in the $\lambda$ sample.
In each panel, the $\lambda$ measurements are plotted from lowest to highest,
with blue points for the $\lambda$ sample and purple points for the
$\psi$ sample.
The left panel shows all available measurements.
The center panel omits data from the gravity darkening
method. The right panel is for the weak-tides
sample.
For each measurement, the $1\sigma$ uncertainty interval is plotted as a vertical line.
}
\label{fig:pseudo_cdf}
\end{figure*}

\subsection{Weak tides}
\label{sec:decoupled}

After applying the weak-tides restriction,
the $\lambda$ sample consists of 70 giant planets for
which only the projected obliquity has been
measured, and the $\psi$ sample consists of
31 giant planets for which
full-obliquity information is available. Because tidal
dissipation is expected to be negligible,
the obliquity distribution inferred for these samples may more closely reflect the outcome of the formation and migration mechanisms.

As in Section~\ref{sec:selection}, the parameters were inferred for 4 different models for the obliquity distribution,
and different choices for the sample and likelihood function.
Table~\ref{tab:par_decoupled} gives the credible intervals of the parameters of the models
for the obliquity distribution.

The results of analyzing the weak-tides sample
were largely consistent
with the results of analyzing the unrestricted sample.
We expected the weak-tides sample to give
a lower value of $q$, the fraction of systems in the well-aligned Rayleigh component, and a larger value of $\sigma_R$, the scale factor of the well-aligned component, but
these differences did not emerge
at a statistically significant level.
Instead, discrepancies arose between the model posteriors inferred from the $\lambda$ and $\psi$ samples.
Relative to the $\psi$ sample, the $\lambda$ sample
favored a larger width for the well-aligned component ($\sigma_R \approx 10^{\circ}$ vs.\ $2^{\circ}$) and a higher fraction of well-aligned planets ($q \approx 0.7$ vs.\ 0.5).

Table~\ref{tab:R_all} gives the Bayes factor of each model (relative to Model 1)
for samples with weak tides.
Regardless of the sample and likelihoods
that were used, the single Rayleigh distribution was strongly disfavored relative to the multi-component
models, and the multi-component models were supported
by nearly equivalent evidence.

\subsection{Strong tides}
\label{sec:coupled}

We also constructed a sample of giant planets
for which tides are expected to be strong, using the
criteria
$M_p > 0.3 M_{\rm Jup}$,
$a / R_{\star} > 7$,
and $T_\mathrm{\rm eff} < 6250$~K.
After applying these restrictions,
the $\psi$ sample has 11 members
and the $\lambda$ sample
has 12 members.
These systems might be expected
to have systematically low obliquities due
to tidal obliquity damping, and indeed, the data are all consistent
with zero obliquity, with varying degrees of uncertainty.

For this sample, the multi-component models are unjustified,
and only Model 0 was fitted.
For the three choices of sample and likelihood ($\psi$, $\lambda$, and $\psi\,\&\,\lambda$),
the Rayleigh scale factor was found
to be $4^{+3}_{-2}$, $5^{+4}_{-2}$, and $4^{+2}_{-2}[^{\circ}]$, respectively. 
The $95$th percentile upper bounds on the Rayleigh scale factor are $9$, $11$, and $7[^{\circ}]$, respectively.

\section{Selection biases}
\label{sec:bias}

The frequentist tests of the $\psi$ sample
assigned a low probability
to an isotropic distribution for the misaligned systems,
and our Bayesian analysis
of the $\psi$ sample
gave somewhat stronger evidence for a $90^\circ$ peak
than an isotropic distribution.
Yet, our analysis of the $\lambda$ sample
favored an isotropic distribution. 
Additionally, the posterior distributions of the model
parameters only partially overlapped between the $\psi$ and $\lambda$ samples. 
Although
the Bayes factors were modest in all cases,
we wondered how and why the $\psi$ sample might differ
from the $\lambda$ sample.
Figure~\ref{fig:pseudo_cdf}
shows the projected-obliquity measurements from the $\lambda$ sample alongside
those from the $\psi$ sample. In this figure,
the $\lambda$ measurements
in the $\psi$ sample appear to be systematically higher
than those in the $\lambda$ sample.

This made us question some premises implicit
in all of the work described above,
namely, that the planets with full-obliquity information
are drawn randomly from the same parent population
as the planets with only projected-obliquity information, and that the availability of
full-obliquity information is independent of the
value of the obliquity.
In this section, we consider the possibility that
these premises are false.
Below, in Section~\ref{sec:bias_obs}, we describe
some statistical tests,
and in Sections~\ref{sec:RM_bias}, \ref{sec:vsini_bias}, and \ref{sec:GD_bias},
we discuss some possible sources of bias
inherent in the measurement techniques.

\begin{figure*}[t]
\gridline{\fig{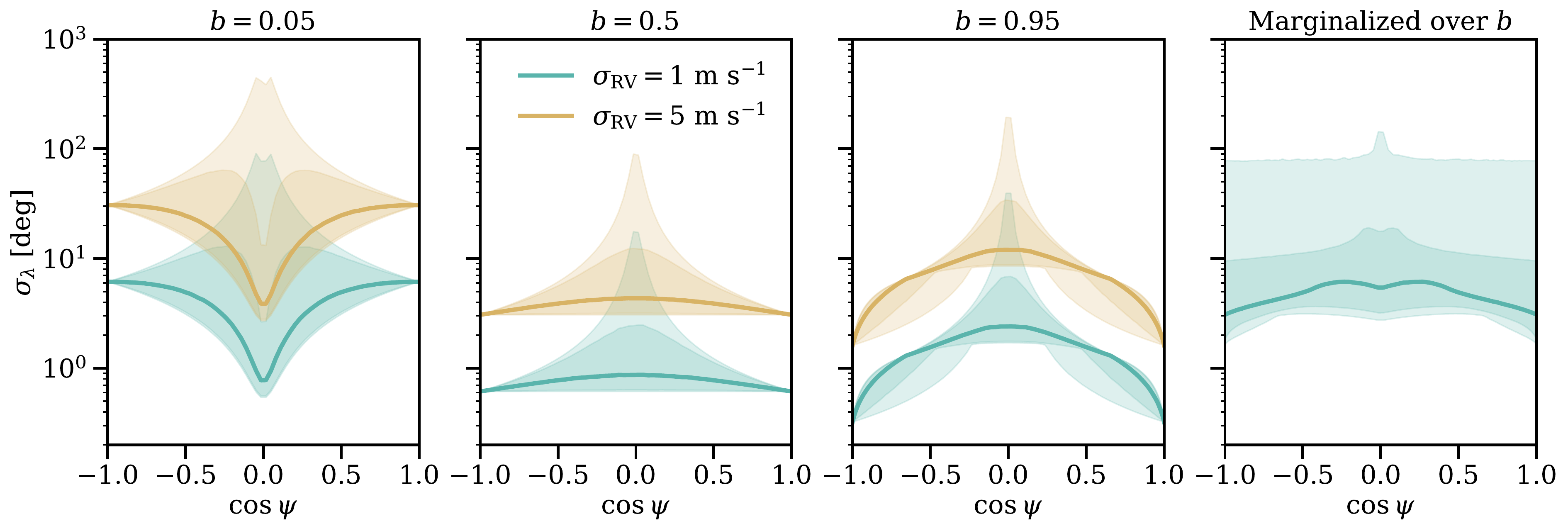}{\textwidth}{} }
\vskip -0.25in
\caption{ The distribution of achievable measurement precision in $\lambda$
for the nominal system described in the text,
assuming random stellar orientations and either a fixed impact parameter or a uniform distribution of impact parameters.
The median value of  $\sigma_\lambda$ is presented as a solid line, and the shaded regions demarcate the $1\sigma$ and $2\sigma$ percentiles. Color indicates the assumed radial velocity measurement uncertainty: $1~\mathrm{m~s}^{-1}$ (blue) or $5~\mathrm{m~s}^{-1}$ (gold). 
 }
\label{fig:RM_selection}
\end{figure*}

\subsection{ Observational evidence }
\label{sec:bias_obs}

To test the null hypothesis that the stellar obliquities
in the $\psi$ and
$\lambda$ samples are drawn from the same distribution,
we applied
the Kolmogorov-Smirnov (KS) test in a series of 1{,}000 Monte
Carlo trials.  For a given trial, we drew (with replacement)
$N_\psi$ values of $\lambda$ 
from the $\psi$ sample, and $N_\lambda$
values of $\lambda$ from the $\lambda$
sample.
These values were drawn from the
measurement posteriors for each system.
Then, the two-sample KS test was applied to compute
the $p$--value for the null hypothesis that the $\lambda$
values are drawn from the same distribution.
The null hypothesis was ``ruled out'' ($p< 5\%$)
in $\lesssim 5\%$ of the trials.
This was true when the method was applied to the entire planet
sample, and when it was applied after enforcing either the weak-tides or strong-tides
restrictions. Two-sample Anderson-Darling tests also ruled out the null hypothesis in  $\lesssim5\%$ of the trials, for each choice of sub-sample.
The median $p$-value was $50\%$ for the whole sample and $40\%$ for the tidally-decoupled sample.
Thus, these tests did not reject
the hypothesis that the $\lambda$ measurements
in the $\psi$ sample are drawn from the same
distribution as the $\lambda$
measurements
in the $\lambda$ sample.

\subsection{ Rossiter-McLaughlin effect }
\label{sec:RM_bias}

For most of the systems in the $\psi$ sample, 
$\lambda$ was measured with the Rossiter-McLaughlin (RM) effect
and $i_\star$ was measured or constrained with a different method.
The RM effect, in brief, is the distortion of a star's spectral
lines that occurs during transits due to stellar rotation.
For example, when a planet blocks a portion of the approaching half of the stellar
disk, the blueshifted components of the spectral
lines are slightly suppressed, which can be detected directly
in the line profile or as an apparent redshift of the entire line.
This ``anomalous'' Doppler shift
depends chiefly on $v \sin i$, $\lambda$, the transit depth $\delta \approx (r/R)^2$, and
the transit impact parameter $b$.
\cite{Gaudi2007} derived an approximate formula, later
corrected by \cite{Albrecht2022}, for the achievable uncertainty 
in $\lambda$ as a function of the system parameters:
\begin{equation}
    \label{eqn:RM_fisher}
    \sigma_\lambda = \frac{ \sigma_{v} /\sqrt{N} }{ v \sin i  } \left ( \frac{r}{R} \right)^{-2} \left [ \frac{ (1-b^2) \cos^2 \lambda + 3 b^2 \sin^2 \lambda }{ b^2(1-b^2) } \right ]^{1/2}.
\end{equation}
The derivation assumes there are $N$ uniformly spaced radial velocity measurements, each with an uncertainty of $\sigma_v$.
Limb darkening is neglected, as are the uncertainties
in $b$ and $r/R$.

The dependence of $\sigma_\lambda$ on $\lambda$ raises
the question of whether the sample of systems with
published obliquity measurements is biased
in favor of obliquities that tend to
allow for smaller values of $\sigma_\lambda$ and consequently
higher probabilities to detect the RM effect.
To understand this source of bias,
we conducted a series of Monte Carlo trials
involving a hypothetical 1\,$R_\odot$ star
with rotation velocity $v=2.5~\mathrm{km~s}^{-1}$
and a transiting Jupiter-sized planet.
In each trial, the star's orientation was drawn
randomly from an isotropic distribution.  Figure~\ref{fig:RM_selection} shows $\sigma_\lambda$ as a function of $\psi$ for three illustrative values of the impact parameter, and for
two choices of $\sigma_v$.
The bias depends strongly on $b$, reflecting
the limiting behaviors of Eqn~\ref{eqn:RM_fisher}:
$\sigma_\lambda \propto \cos^2 \lambda$ as $b\rightarrow 0$,
and $\sigma_\lambda \propto \sin^2 \lambda$ as $b\rightarrow 1$.
The rightmost panel of Figure~\ref{fig:RM_selection} shows
that after
marginalizing over a uniform distribution of $b$
between 0 and 1,
$\sigma_\lambda$ varies weakly with $\psi$.
Values of $\psi=0^\circ$ and $180^\circ$ lead
to average values of $\sigma_\lambda$ that are about two times
lower than when $\psi=90^\circ$.
Thus, we expect that any such bias in the real sample
to be small and to
favor well-aligned and anti-aligned systems
rather than producing a peak at $\psi=90^\circ$.

\begin{figure*}[t]
\gridline{\fig{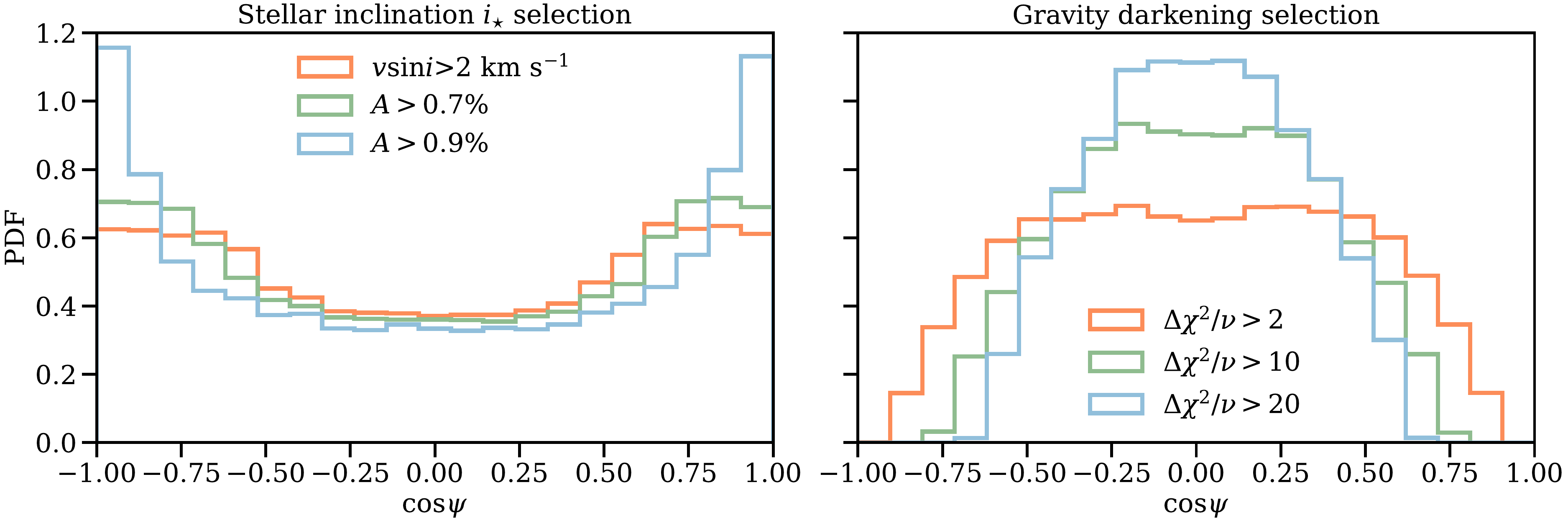}{\textwidth}{} }
\vskip -0.25in
\caption{ Biases in the $v\sin i$ method (Section~\ref{sec:vsini_bias}) and the gravity darkening method (Section~\ref{sec:GD_bias}) shape the recovered obliquity distribution.
Each panel shows the obliquity distribution of hypothetical stars drawn from
a randomly-oriented population, after applying 
increasingly stringent detection thresholds.
\textit{Left}: Obliquities measured with the $v\sin i$ method, using
three detection thresholds: (i) $v \sin i > 2~\mathrm{km~s}^{-1}$, (ii) photometric amplitude $A > 0.7\%$, where $A = f_\mathrm{spot} \sin i$, and (iii) $A > 0.9\%$. The $v\sin i$ method favors $\psi=0^\circ$ and $180^\circ$.
\textit{Right}: Obliquities measured with the gravity-darkening method. The detection thresholds were parameterized by the difference in reduced-$\chi^2$ between modeling the light curve with gravity-darkening or a standard transit profile. }
\label{fig:vsini_and_gd}
\end{figure*}

\subsection{The $v\sin i$ method}
\label{sec:vsini_bias}

Many of the constraints on stellar inclination
come from the combination
of measurements of the star's projected
rotation velocity $v \sin i$,
radius $R$,
and rotation period $P_\mathrm{\rm rot}$, based on Equation~\ref{eq:vsini}.
Any selection biases regarding $R$ are unlikely to depend
on the star's orientation, but
$v\sin i$ and $P_{\rm rot}$ are potentially problematic.
Measurements of $v\sin i$
tend to be limited to an accuracy of 1--2~km~s$^{-1}$
by systematic effects; thus, we would expect
the fractional uncertainty in
$v\sin i$ to be larger
when $\sin i$ is small.
Small values of $\sin i$ also make
measurements of $P_{\rm rot}$ more difficult,
because the photometric
variability due starspots and plages
has a lower amplitude at low inclinations.

We conducted more Monte Carlo trials
to understand the consequent selection biases
as a function of $\psi$.
The intrinsic stellar properties were held
fixed: $R=R_{\odot}$, $P_\mathrm{\rm rot}=15$~days,
and a maximum loss of light due to spots
of $f_{\rm spot}=1\%$.
In each trial,
the star's orientation and the planet's obliquity were
drawn independently from an isotropic distribution. 
We assumed the photometric variability
amplitude $A$ to be $f_{\rm spot}\sin i$, an approximation that assumes the spots are small
and nearly equatorial, and neglects the effects
of limb darkening \citep[see, e.g.,][]{JacksonJeffries2012,Mazeh2015}.
The stellar inclination was considered ``measurable''
when $v \sin i > 2~\mathrm{km~s}^{-1}$ and
$A>A_{\rm min}$, or
equivalently,
\begin{equation}
    \sin i > \max \left [ \frac{2~\mathrm{km~s}^{-1}}{v},  \frac{ A_{\min} }{ f_\mathrm{spot} } \right ].
\end{equation}
For the adopted stellar parameters,
the $v \sin i$
criterion is the limiting factor
when $A_{\min} \lesssim 0.6\%$.
Figure~\ref{fig:vsini_and_gd} shows the obliquity distribution of stars with measurable inclinations,
for increasingly stringent detection criteria.

These selection effects would cause the sample
to be biased in favor of well-aligned
and anti-aligned systems, for which
the inclination is guaranteed to be close
to $90^\circ$.  Misaligned systems, and perpendicular
systems in particular, would occasionally
be observed pole-on and are thereby disfavored.

\subsection{ Gravity Darkening  }
\label{sec:GD_bias}

The photosphere of a rapidly rotating star
is cooler and fainter near the equator than the poles,
because the equatorial region is centrifugally lifted to
higher elevation.
This effect, known as gravity darkening,
causes distortions of a transit
light curve that depend on both $\lambda$ and $i_\star$ in addition
to the usual transit parameters \citep{Barnes2009}.
For five of the systems in the $\psi$ sample
(KELT-9, KELT-17, Kepler-13, MASCARA-4, and WASP-189),
the stellar obliquity was determined by
modeling transit light curves affected by
gravity darkening. In contrast, none of
the measurements in the $\lambda$
sample relied on gravity darkening.
Could this difference account for the differences
in the inferred obliquity distributions of the
$\psi$ and $\lambda$ samples?

Gravity-darkening anomalies in the light curve are typically on the order of $\lesssim$0.1\%,
making them challenging to detect,
and the
detectability depends on the stellar orientation.
When $\psi$ is near $0^\circ$ or $180^\circ$,
the transit light curve retains its usual ``U-shape'' with mirror
symmetry around the time of minimum light, causing
the parameters describing gravity darkening to be
strongly covariant with the usual transit
parameters.
For other obliquities, 
the transit light curve develops an
asymmetry or a ``W-shape'',
either of which reduces the covariance between
the gravity-darkening parameters and the other
parameters, allowing the gravity-darkening
signal to be detected and interpreted
with greater confidence.
Thus, the sample of
systems for which the gravity darkening
signal has been reported is likely to be biased
against well-aligned and anti-aligned systems.

To understand the extent of this bias,
we performed another round of Monte Carlo trials
involving a single star that can have any
obliquity and a transiting planet
that can have any impact parameter.
Based on the properties of one of the best
studied systems, Kepler-13A \citep{Masuda2015},
we fixed the properties of the hypothetical
star to be
$M=1.80\,M_{\odot}$,
$R=1.71\,R_{\odot}$,
$P_\mathrm{\rm rot} = 20$~hours,
$r/R =0.085$, and $P_{\rm orb} =1.76$~days. 
A quadratic limb-darkening law was
adopted, with coefficients given by
\cite{Masuda2015}: $u_1 + u_2 = 0.53$ and $u_1 - u_2 = 0.26$.
The effective temperature $T$ was assumed to vary with surface gravity $g$ as 
\begin{equation}
    T = (7500\,{\rm K}) \left ( \frac{g}{g_\mathrm{pole}} \right )^{1/4}.
\end{equation}
In each of $2 \times 10^5$ trials,
the stellar orientation was drawn from an isotropic distribution, and the planet's transit
impact parameter was drawn from 
a uniform distribution between 0 and 1.
We generated a
transit light curve
exhibiting the effects of gravity darkening
using the \texttt{pytransit} code
written by \cite{Parviainen2015}.
We calculated the difference in reduced-$\chi^2$ between the best-fit model
including gravity darkening and the best-fit
model with no gravity darkening.
For computational efficiency, only $i_{\star}$, $\lambda$, $P_\mathrm{rot}$, the limb-darkening coefficients, and $r/R$ were treated as free parameters. 
Through these Monte Carlo trials, we mapped the relation between $\Delta \chi^2 / \nu$ and $\psi$.

Figure~\ref{fig:vsini_and_gd} shows
the obliquity distribution of those
systems for which $\Delta \chi^2 / \nu$ exceeds
a threshold value. Results for three different
choices of the threshold value are shown.
Even for modest detection thresholds (i.e., $\Delta \chi^2 / \nu>2$), the obliquity distribution of the systems with detectable gravity-darkening
signals is biased strongly against
$\psi=0^\circ$ and $180^\circ$.

The bias goes in the right direction
to explain why the $\psi$ sample
shows stronger evidence for a ``pile-up'' of
obliquities near $90^\circ$ than
the $\lambda$ sample.
To pursue this point,
we repeated the modeling
procedure described in
Section~\ref{sec:selection} after
excluding the 5 gravity-darkening
systems
from the sample.
The resulting Bayes factors, relative to Model 1, are presented in the third row of Table~\ref{tab:R_all}.
After omitting the gravity-darkening measurements, the $\psi$ sample shows a weaker preference for a peak
at 90$^\circ$ in the obliquity distribution.
This trend is evident in Figure~\ref{fig:pseudo_cdf}.

\begin{figure*}[t]
\gridline{\fig{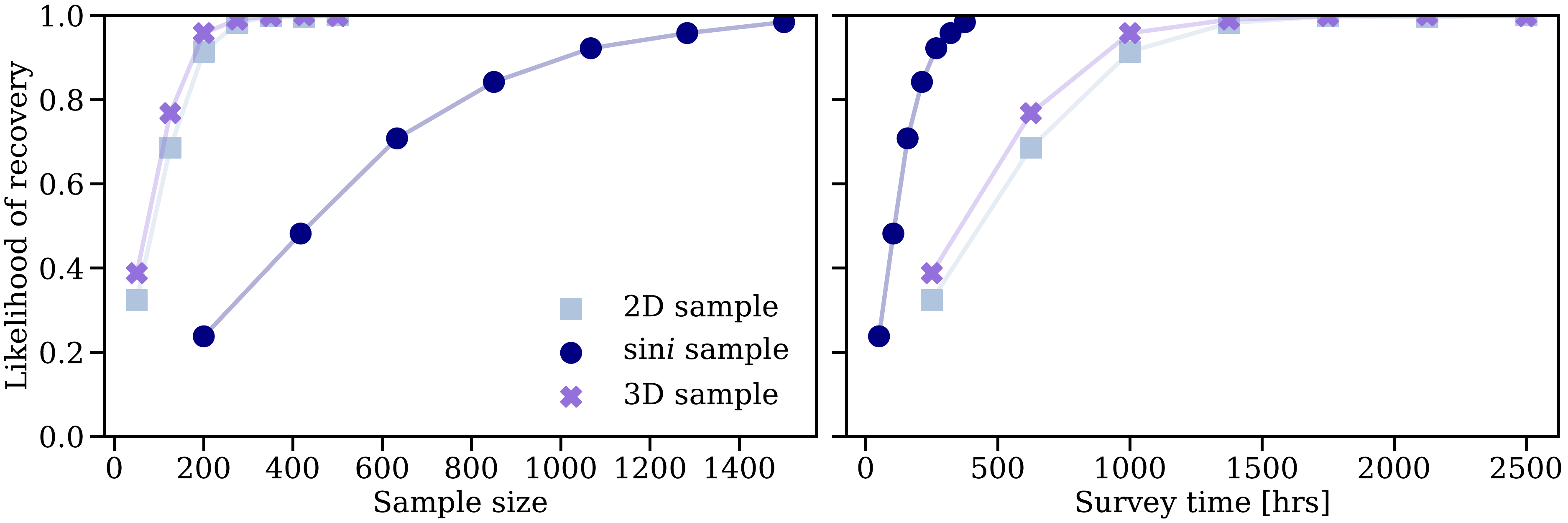}{\textwidth}{} }
\vskip -0.25in
\caption{ Observational investment required to recover a peaked
obliquity distribution, for three idealized campaigns:
(i) measuring $\lambda$ only,
(ii) measuring $\sin i$ only,
and (iii) measuring both $\lambda$ and $\sin i$.
The vertical axis is the fraction of realizations in which the
postulated peak (see text) is successfully identified. 
In the left panel, the horizontal axis is the sample size;
in the right panel, the horizontal axis is the total observing time, assuming $5$~hr per $\lambda$ measurement and $0.25$~hr per $v \sin i$ measurement. 
}
\label{fig:pop_syn}
\end{figure*}

\section{Population Synthesis}
\label{sec:pop}

Bayesian analysis of the $\psi$ and $\lambda$ samples did not distinguish between a peaked or flat distribution for the misaligned systems.
While this result held for both the $\psi$ and $\lambda$ samples---as well as the weak-tides sub-sample---minor discrepancies persist:
the $\psi$ sample marginally favors a $90^\circ$ peak, more so than the $\lambda$ sample, and the posteriors of the model parameters only partially overlap between the $\psi$ and $\lambda$ samples.
Prompted by these mysteries, we investigated selection biases in Section~\ref{sec:bias}. 

However, the impacts of selection biases are difficult to constrain for heterogeneous samples. Stars with
different characteristics may have different stellar
obliquity distributions, and they may
also have different characteristics that affect whether
the stellar obliquity can be measured.
For example, the ability to measure rotation periods from photometry is easier for cool spotted stars with convective envelopes than
for hotter stars with radiative envelopes, and there is also
evidence that cool and hot stars have intrinsically
different
obliquity distributions.
We advocate for the creation of a more homogeneous sample to understand and overcome these biases.
To achieve this goal, would it be better to
focus on $\lambda$ measurements, which are observationally
expensive, or on $\sin i$ measurements, which bear less
information but are less expensive?

We performed some Monte Carlo experiments to begin answering
this question. We generated a synthetic sample of obliquities drawn from an isonormal model with $w=0.7$, a level of ``peakedness" motivated by the fits to the current sample.
Host star properties and measurement uncertainties
were assigned for each draw by randomly selecting host stars and their
measurement uncertainties from the observed sample.
This process neglected the dependence of $\sigma_\lambda$ on $\lambda$ and $\sigma_{v \sin i}$ on $v \sin i$, but retained the $v \sin i$ selection bias by enforcing realistic uncertainties $\sigma_{v \sin i}$.
For a given synthetic sample, we computed the Bayes factor
comparing the isonormal model and an isotropic model.
The peaked distribution was deemed ``recovered'' if the Bayes factor for the isonormal model relative to the isotropic model was greater than $10$.
When calculating Bayes factors,
we considered three scenarios:
(i) measuring $\lambda$ only, via the RM effect;
(ii) measuring $\sin i$ only, by combining $v \sin i$, $R_\star$, and $P_\mathrm{rot}$;
and (iii) measuring both $\lambda$ and $\sin i$ (the $\psi$ sample).
We did not consider gravity darkening measurements because
of the strong bias of that method against aligned and anti-aligned stars.
For a given choice of sample size and measurement technique,
500 trials were conducted.

Figure~\ref{fig:pop_syn} shows the fraction of trials in
which the peaked model was successfully recovered.
To reach a success rate of $>90\%$,
the necessary sample size was 180, 200, and 1000, for the $\psi$, $\lambda$, and $\sin i$ samples, respectively.
As expected from the results presented in Section~\ref{sec:obliquity_only}, a single
$\lambda$ measurement is typically more valuable than a single
$\sin i$ measurement.  
However, a typical RM measurement requires a
time-critical observation lasting $\approx\,5$\,hr on
a large telescope with a stabilized high-resolution
spectrograph. In contrast, measuring $v \sin i$ with such a
telescope requires no more than 15 minutes (including overheads), does not require a highly stable wavelength calibration, 
and can be conducted at any time.
Applying these fiducial costs of 5\,hr per $\lambda$ measurement
and $0.25$\,hr per $\sin i$ measurement,
we found that the necessary observing time is 990, 980, and 250~hr for the $\psi$, $\lambda$, and $\sin i$ samples, respectively.
Thus, in these simulations, the $v \sin i$ technique is favored,
even without accounting for bad weather and scheduling constraints
that would also favor the $v\sin i$ technique. Because of
all the simplifying assumptions, these
results should be considered illustrative rather than
quantitatively reliable.

\section{Summary and Conclusions}
\label{sec:concl}

Even before the first misaligned planetary systems
were discovered, theorists explored mechanisms
for tilting a planet's orbital plane
away from the star's equatorial plane \citep[e.g.,][]{YuTremaine2001,Fabrycky2007}.
Theoretical activity accelerated after
the first few sightings of grossly
misaligned systems
\citep[e.g.,][]{Hebrard2008, Winn2009, Queloz2010}, leaving
us with numerous proposed mechanisms for
spin-orbit misalignment, such as magnetic warping
of the protoplanetary disk \citep{Foucart2011}, Von Zeipel–Kozai–Lidov cycles \citep{Fabrycky2007, Naoz2011}, planet-planet scattering \citep{Chatterjee2008}, and inclination excitation from secular resonance crossing \citep{Petrovich2020}.
Characterizing the observed obliquity distribution is an
important goal in testing these theories for spin-orbit
misalignment.

We applied hierarchical Bayesian inference to a sample of 174 planets (72 with full-obliquity constraints and 102 with only projected-obliquity constraints). Through Bayesian
model selection, we tried to establish: (i) Is the obliquity
distribution compatible with a single smooth distribution
centered on $0^\circ$? (ii) Are the misaligned systems consistent with isotropy or do they exhibit a concentration? (iii) If so, what is the preferred value of the obliquity?

A single Rayleigh distribution was found to be a terrible
description of the data, relative to the models
in which some systems are drawn from a Rayleigh
distribution and others are drawn from a broader
distribution. 
The sample with full-obliquity constraints showed greater evidence for a peak in the obliquity distribution
at about $90^\circ$, relative to a model in which
the misaligned systems are drawn from an isotropic
distribution. However, the odds ratios were modest
($\lesssim$3), and the results are likely affected
by the selection bias of the gravity-darkening method
that favors near-perpendicular systems.
Analyzing the larger sample of systems
with only projected-obliquity constraints, or the union
of all systems with any kind of obliquity constraints,
the Bayesian evidence
did not reveal
any preference for a concentration near 90$^\circ$.
Our investigation of selection biases showed
that the gravity-darkening method favors systems
with large obliquities, as noted above, while
the other methods favor aligned and anti-aligned systems.

We believe that major progress on measuring the obliquity
distribution will require the construction of a large
sample of systems that have been measured with the same
technique (or combination of techniques), coupled with a good
understanding of the selection function.
It is not clear how to achieve this goal.
The RM technique is observationally expensive, typically
requiring 4-5 hours per system using a large telescope
equipped with a high-resolution spectrograph.
The gravity-darkening technique requires transiting
planets around rapidly
rotating stars, of which few are known; it also
requires very precise transit photometry.
The $v\sin i$ technique is the only one for which
it seems feasible to expand the sample to thousands
of stars or more. Only a single high-resolution spectrum
of each system is needed, along with the stellar rotation
period, which can be obtained from
ongoing or planned wide-field photometric time-domain surveys.
The ideal types of star are probably early G stars
or late F stars, for which
moderately rapid rotation is expected
($\gtrsim$\,5~km/s)
and for which rotationally-induced photometric variations
are also expected. 
Our population-synthesis calculations demonstrated that a $v \sin i$ survey would require less observing time than a RM survey
(by a factor of a few) to distinguish between a peaked and isotropic obliquity distribution, although these calculations
were based on many simplifying assumptions. Before deciding
on an observing program, other considerations
would need to be taken into account,
such as the availability and reliability of photometric rotation periods,
and the limited number of transiting
planets that are known to exist around early G and late F stars.
Nobody said that solving the spin-orbit misalignment
problem would
be easy. 

\acknowledgments
We thank Jiayin Dong and Dan Foreman-Mackey for helpful conversations and for sharing their preprint of a paper
about modeling the obliquity distribution.
JS acknowledges support by the National Science Foundation Graduate Research Fellowship Program under Grant DGE-2039656. 
SHA acknowledges the hospitality of the Department of Astrophysical Sciences at Princeton University and support from the Danish Council for Independent Research through a grant, No.\,2032-00230B.
Any opinions, findings, and conclusions or recommendations expressed in this material are those of the author(s) and do not necessarily reflect the views of the National Science Foundation.

\bibliography{paper}%

\end{document}